\documentclass[journal]{IEEEtran}
\usepackage{cite}
\ifCLASSINFOpdf
  \usepackage[pdftex]{graphicx}
\else
\fi
\usepackage{amsmath}
\usepackage{array}
\usepackage{url}
\usepackage{booktabs}
\usepackage{amssymb}
\usepackage{hyperref}

\begin{document}
\title{BERT Embeddings Can Track Context in Conversational Search}

\author{Rafael~Ferreira,
        David~Semedo,
        and~João Magalhães% <-this % stops a space 
        \\ NOVA School of Science and Technology, Portugal \\
        rah.ferreira@campus.fct.unl.pt, \{df.semedo, jmag\}@fct.unl.pt}
%\thanks{Rafael Ferreira, David Semedo and João Magalhães. Shell was with the Department of Electrical and Computer Engineering, Georgia Institute of Technology, Atlanta,


% TODO change this in final version
\markboth{Preprint}%
{R. Ferreira \textit{et al.} Preprint: BERT Embeddings Can Track Context in Conversational Search}



\maketitle

\begin{abstract}
The use of conversational assistants to search for information is becoming increasingly more popular among the general public, pushing the research towards more advanced and sophisticated techniques.
In the last few years, in particular, the interest in conversational search is increasing, not only because of the generalization of conversational assistants but also because conversational search is a step forward in allowing a more natural interaction with the system. 

In this work, the focus is on exploring the context present of the conversation via the historical utterances and respective embeddings with the aim of developing a conversational search system that helps people search for information in a natural way. In particular, this system must be able to understand the context where the question is posed, tracking the current state of the conversation and detecting mentions to previous questions and answers. We achieve this by using a context-tracking component based on neural query-rewriting models. Another crucial aspect of the system is to provide the most relevant answers given the question and the conversational history. To achieve this objective, we used a Transformer-based re-ranking method and expanded this architecture to use the conversational context.

The results obtained with the system developed showed the advantages of using the context present in the natural language utterances and in the neural embeddings generated throughout the conversation.
%TODO
%The models developed also achieved results superior to the baseline model in the TREC Conversational Assistance Track (CAsT) 2020.
\end{abstract}


% \fi
\IEEEpeerreviewmaketitle

\section{Introduction}
\IEEEPARstart{C}{onversational Search} 
%Conversational search 
systems are an emerging research topic and a step forward from traditional search engines, allowing a more natural interaction with an intelligent agent.
The potential of these kinds of systems is very high, with applications to various domains, such as e-commerce, medical question answering, and others. Conversational assistants, such as Siri, Alexa, and Bixby, have been around for a few years, with substantial improvements since their creation. However, their ability to support conversational search is up to this time still limited. 
Hence, this a good opportunity to address this limitation with stronger context-tracking models that understand the underlying context of the conversation in order to retrieve relevant information.

This work is focused on search and answer selection (QA) in a conversational search setting. 
This more challenging scenario departs from typical single-turn QA and search systems, where the system only needs to answer the current question of the user without any notion of previous questions/answers.
In particular, we highlight the use of state and context from previous questions and answers.

The formal definition of conversational search provided in \cite{castoverview} is: given a series of natural language utterances ($u$) about a given topic \textit{T}: \begin{equation}
    T=\{u_1, \ldots u_i, \ldots u_n\},
\end{equation}
the task is to identify relevant passages $P_i$ for each turn (user utterance) $u_i$ that satisfies the information needs in round $i$ given the conversation context, i.e., the previous conversation turns $u_1,\ldots, u_{i-1}$. 

As stated, the conversational context assumes a central role, therefore, we have to not only manage the current query but also have a notion of the context where the query is posed. 
We believe that the context can be seen in both previous queries and answers, so the objective of this work is to research methods that can improve the search results by tracking the context present in: 
\begin{itemize}
    \item The conversation's utterances in natural language;
    \item The neural embeddings generated by a conversational neural-state tracker.
\end{itemize} 

So to leverage the importance of the state of the conversation in searching for the correct information, we propose a three-stage architecture for conversational search composed of: (i) a conversational query rewriting, (ii) retrieval, and (iii) context-aware re-ranking components that use the context present in the previous queries and answers, and in the embeddings generated by a Transformer re-ranking model.

Hence the core contributions of this work are the following: 
\begin{itemize}
    \item An investigation of how the conversational context can be explored as a query rewriting problem and as a neural state-tracking problem;
    \item We propose two new architectures based on RNNs~\cite{gruOriginal, Hochreiter1997LongSM} and Memory Networks~\cite{sukhbaatar2015endtoend} that bring state-of-the-art neural re-ranking models to the conversational search domain;
    \item A systematic experimental evaluation of each component developed that shows the advantages of tracking the context in the historical utterances and in neural embeddings.
\end{itemize}

In the following section, we discuss the related work. In section \ref{work_section}, we detail the conversational search pipeline: the conversational query rewriting and the re-ranker. Evaluation is performed in Section \ref{evaluation_section}, and Section \ref{conclusions_section} presents the conclusions of this work.

\section{Related Work}
In this chapter, we present the concepts related to conversational search and the current methodologies and algorithms used in the various components of such systems.

\subsection{Conversational Query Rewriting}
%\textbf{Conversational Query Rewriting}. 
To address the omissions (coreferences) present in a conversational query, we can perform query rewriting to obtain context-independent queries that can be used in typical IR systems. 

An example of this is showed in~\cite{canYouUnpackIt}, where researchers created the CANARD dataset by manually rewriting the queries in QuAC~\cite{quac_dataset} (a conversational QA dataset about a single passage) to train a model in the conversational query rewriting task. In particular, a sequence-to-sequence model with an attention and copy mechanism is fed with the full conversation history and the query to be rewritten. In~\cite{query_rewriting_limited_supervision} a similar input structure is used to fine-tune a BERT model on a binary term classification task that aims to add to the current query terms from previous turns. In~\cite{t5conversational} is employed a pre-trained T5 model~\cite{t5} fine-tuned on the CANARD dataset~\cite{canYouUnpackIt} to construct the context-independent query, and achieved state-of-the-art performance on that query-rewriting task. 
Finally, in \cite{gpt2_query_rewriting} is presented a few-shot generative method based on a GPT-2 model fine-tuned on a large amount of weakly supervised data.

\subsection{Neural Passage Ranking}
%\textbf{Neural Passage Ranking}. 
Some attempts to use neural models for the re-ranking task were defined by the use of pre-computed word embeddings \cite{word2vec} that would then be used to calculate the similarity between the words in the query and document, or to calculate a measure of relevance \cite{drmm, knrm, convknrm}. 

After the introduction of the large pre-trained language models based on the Transformer~\cite{vaswani2017attention} architecture such as BERT~\cite{bertOriginal} and others \cite{roberta,t5,yang2019xlnet}, the focus shifted to use the contextual embeddings generated by these models to assess the relevance of a document to a query by their ability to model more complex interactions between the text in both sources~\cite{passagererankingbert, t5_reranking}.
In particular, in \cite{passagererankingbert}, a BERT model is fine-tuned on a binary relevance classification task using the MS MARCO dataset~\cite{marcoDataset} and the cross-entropy loss function with the input being a query and a relevant or non-relevant passage.
In a more recent work \cite{t5_reranking} the same dataset is used but the BERT model is replaced by the encoder-decoder model T5 \cite{t5}, being able to surpass the equivalent BERT method.

\subsection{Conversational State-Tracking} 
%\textbf{Conversational State-Tracking}. 
The state tracking task is one of the most important aspects of conversational systems because it gives context to the current conversation turn. 

In \cite{hredConversational} is proposed the Hierarchical Recurrent Encoder-Decoder (HRED) model. In this model, a conversation is seen as a sequence of tokens, and a sequence of utterances that are encoded at different levels. So in a conversation, the encoder RNN maps each utterance to an utterance vector representing the hidden state. The RNN at a higher-level keeps track of the past utterances (context) by processing each utterance vector iteratively. The next utterance prediction is performed by a decoder RNN, which takes the hidden state of the context RNN and produces a probability distribution over the tokens in the next utterance.

Because of the small memory and gradient vanishing problems of RNNs, in \cite{weston2014memory} is presented a class of models called memory networks. Memory networks address the RNNs limitations by providing an easy way to read and write part of a long-term memory component. This model utilizes an explicit memory to store previous representations, being able, in each turn, to account for the full conversational history independently, instead of relying on a single state like the HRED model \cite{hredConversational}. 
In \cite{sukhbaatar2015endtoend}, is presented an extension to the original memory networks \cite{weston2014memory} which uses an end-to-end architecture with a recurrent attention mechanism to read the memory that can be successfully trained via back-propagation.
Other applications of memory networks to model conversational context include Visual Dialog~\cite{visual_dialog}, which uses both images and text, and the Wizard of Wikipedia~\cite{wizard_wikipedia}. In Visual dialog, a memory network is used to keep track of previous question-answer embeddings generated by an LSTM about an image.
In Wizard of Wikipedia, memory networks are used in conjunction with BERT embeddings to maintain the conversational context, with the objective of simulating a conversation.

\subsection{Conversational Search Systems}
%\textbf{Conversational Search Systems}. 
Open-domain conversational search systems are a particular type of information retrieval systems where the context of the conversation is of great importance to answer the user’s queries. 
This topic has been studied in~\cite{evaluation_conversational_assistants} that showed that users are willing to utilize conversational assistants as long as their needs are met with success. 
The interest in these systems has been on the rise with the generalization of intelligent conversation assistants (e.g., Alexa, SIRI), but there were not any dedicated datasets for conversation search until TREC CAsT~\cite{castoverview}. With the creation of this focused dataset, the submissions were diverse, including traditional retrieval-based methods, neural models, and knowledge enhanced methods.

In particular, the top-performing method \textit{HistoricalQE}~\cite{historicalqe} uses contextual query rewriting. This method uses the scores of the retrieval model for each word to identify query and session words to expand the query. The re-ranking part is performed by a BERT model~\cite{bertOriginal} fine-tuned on a binary query-passage relevance estimation task~\cite{passagererankingbert}. Other works include \textit{ClacBase}~\cite{WaterlooClarke} that uses a fine-tuned BM25 model with pseudo-relevance feedback, in conjunction with the AllenNLP~\cite{AllenNLP} coreference resolution model to give context to the queries. In~\cite{predicting_relevant_turns} is presented a system that also uses AllenNLP for coreference resolution while also adding to each conversational query, relevant utterances selected by a fine-tuned BERT model. After this is also employed another fine-tuned  BERT model to perform the re-ranking of the passages.

%\IEEEpubidadjcol

\section{Tracking the Conversational Context}
\label{work_section}
As stated before, in this work, we explore the context present in the natural language utterances and in the neural embeddings.

Regarding the natural language utterances, we explore query rewriting methods that use the queries from previous turns to rewrite the current conversational query.

In the exploration of the neural embeddings, we generate embeddings with BERT~\cite{bertOriginal} for all turns, and then we use these embeddings to model the context using two architectures:
\begin{itemize}
    \item \textbf{ConvBERT RNN} - An RNN is used to model the conversational context;
    \item \textbf{ConvBERT MemNet} - Memory Networks are used to model the conversational context.
\end{itemize}

\subsection{Conversational Query Rewriting}
\label{sub:text-to-text-model}
As a consequence of the conversational properties of this work, the query can be in a format that doesn't have all of the information needed to answer the user's query. Therefore, it is necessary to have a method of rewriting the query to a non-conversational format.  

Table ~\ref{table:example_conversation_montreal} shows an example of the need for a query rewriting component.
In turn 2, the system needs to understand that ``one'' refers to ``physician's assistant'' (explicit coreference). Another even more challenging example is presented in turn 3 since the ``starting salary'' should be for the ``physician's assistant'' position, despite no direct evidence in the current query text (implicit coreference).
In a regular information retrieval system, it is almost impossible to retrieve relevant information for turns 2 and 3 of the conversation because there is no notion of the context of the conversation. 

\begin{table*}[t]
\centering
\caption{Conversation example about a specific topic.}
\label{table:example_conversation_montreal}
%\resizebox{\linewidth}{!}{%
\begin{tabular}{cll}%{cp{0.5\linewidth}p{0.6\linewidth}}
\toprule
\textbf{Turn} & \textbf{Conversational Query}                & \textbf{Non-conversational Query}       \\ \midrule
1             & What is a \underline{physician’s assistant}? & What is a \underline{physician’s assistant}?           \\
2             & How can you become \underline{one}?    & How can you become a \underline{physician’s assistant}?         \\
3             &  What’s the starting salary in Canada? \ & What’s the starting salary in Canada \underline{for a physician’s assistant}? \\ \bottomrule
\end{tabular}%
%}
\end{table*}

\subsubsection{\textbf{Using Previous Queries.}}
\label{sub_sub_previous_queries}
In conversational search, the incorporation of previous queries can be seen as a simple way of giving context to the current query. If the conversation follows the typical information-seeking pattern of first exploring a general concept and then following this with a more detailed search about that concept or a related one, we can use the previously mentioned terms to expand the query. 
With this in mind, we developed two methods:
\begin{itemize}
    \item \textbf{Prefixing} - Inspired by \cite{WaterlooClarke}, we prefix the first query issued by the user to the current query. We use the first query since it is non-conversational, and it is usually the starting point for the subsequent queries. One of the problems of this approach is that in the event of a topic shift during the conversation, the first query may introduce noise.
    \item \textbf{Union} - Performs the union of the current query with each of the previous queries separately, performing \textit{n} queries (that are easily parallelizable) depending on the turn depth $t$ following equation~\ref{eq_union}:
        \begin{equation}
    n = \begin{cases} 1, & \ if \ \  t\leq2 \\ t-1, & \ if \ t > 2     \end{cases} \ \ \ \ t \geq 1,    \label{eq_union}
    \end{equation}
    After this, we make a union of the results of all the queries and keep the top \textit{k} documents retrieved with the highest retrieval scores, making this a fusion algorithm. With this approach, we expect to obtain the most relevant documents for each combination of queries, removing the documents that appear lower on the rank for each combination of queries.
\end{itemize}
    
\subsubsection{\textbf{Coreference Resolution.}}
\label{sub_sub_allennlp}
Coreference resolution is applied before performing any expansion on the query or search in the dataset. We aim to resolve the coreferences in order to search for the correct information.
To achieve this, we used the toolkit provided by AllenNLP~\cite{AllenNLP}. In specific, we utilized the coreference resolution module that uses the implementation from \cite{allen_coref_resolution} which replaces the GloVe embeddings with BERT embeddings.

In the AllenNLP~\cite{AllenNLP} toolkit, the coreference resolution algorithm receives sentences as input, and outputs a list of mentions that indicate which spans in the input belong to the same mention. 
With the notion that the first mention is the most descriptive of all, we replace all the spans with the same mention by the first one, however, after analyzing the results of this approach we noticed that the algorithm sometimes assigns the same mention to concepts that are not related (coreferent). To mitigate the problem identified in the previous approach, we developed a method called \textbf{Coref-Pronoun} that only replaces the mention if it is needed, i.e., when pronouns are present in the mention. Using this approach, mentions that give some information about the subject remain unchanged.
In our experiments, we saw good results in explicit coreference resolution (Table \ref{table:example_conversation_montreal} Turn 2) but the model lacks the ability to perform context resolution (Table \ref{table:example_conversation_montreal} Turn 3).

\subsubsection{\textbf{Text-To-Text Transfer Transformer (T5).}}
\label{sub_sub_t5}
In our work, we use the characteristics of T5~\cite{t5} to rewrite the queries in a conversational format to form non-conversational queries. To achieve this, the model must be capable of resolving the coreferences and provide the necessary context to queries that are not specific enough. 

To train the T5 model it is necessary to provide an input sequence and a target sequence given as strings. Consequently, in a conversational query rewriting scenario, the input is a conversation, and the target is the last query rewritten.
To accomplish this, our approach follows closely the one presented in ~\cite{t5conversational}, although we use a different input structure. 
In specific, considering $i$ the current turn, $q$ a query, $p$ a retrieved passage, and \textit{[CTX]} and \textit{[TURN]} special separator tokens the input format is defined as:
\begin{equation}
        ``q_i \ [CTX] \ q_1 \ p_1 \ [TURN] \ q_2 \ p_2 \ [TURN] \ \ldots \ q_{i-1} \ p_{i-1}",
\end{equation}
where \textit{[CTX]} separates the current query from the context (previous queries and passages), and \textit{[TURN]} is used to separate the historical turns.

In our initial experiments, we observed that the T5 model was capable of performing explicit coreference resolution like AllenNLP's model (Section \ref{sub_sub_allennlp}). But the main advantage of this model is its ability to perform implicit coreference resolution (Turn 3 of Table \ref{table:example_conversation_montreal}), which is very important in a conversational search system.

% Examples of the inputs, targets, and predicted queries are presented in table~\ref{table:t5example_2}. The first two examples are from CANARD, and the last two are from TREC CAsT~\cite{trecCast}, the conversational search dataset that we introduced in Section \ref{sub:sub:retrieval_datasets}, and that we will evaluate in more detail in Section~\ref{sec:evaluation_dataset}. The way the input is created only affects the CANARD dataset because in TREC CAsT 2019 the historical utterances don’t depend on the responses of the system. 
% As we can see, T5 is capable of resolving ambiguous queries in most situations, however, it sometimes mistakes similar entities when multiple are involved, as evidenced in~\cite{canYouUnpackIt} and in table~\ref{table:t5example_2} example 3, where the model predicts ``throat cancer'' instead of ``lung cancer''. 
% We can also note that this model is more robust than simple coreference resolution, as we can see in example 4, where it includes the words ``Bronze Age Collapse'', even though there is no particular pronoun or coreference mention (implicit coreference). 

% Comparing the T5 model to the previously described AllenNLP coreference resolution model, we can see that both are able to resolve explicit coreferences, i.e., when a particular pronoun or name is used to refer to a previously mentioned entity. The main advantage of this trained T5 model over AllenNLP's model is the ability to perform implicit coreference and context resolution, which is very important in a conversational search system.

\subsection{ConvBERT}
\label{conv_bert}
With the new pre-trained neural language models, such as BERT~\cite{bertOriginal}, it is possible to generate contextual embeddings for a sentence and each of its terms. These embeddings can then be used as input to a model to perform re-ranking of the passages. 
This ranking is usually considered better than the one given by term-matching models because BERT captures the context of the terms in the query and passage, as well as their interactions, being able to judge more thoroughly if a passage is indeed relevant to a query, not by term matching or frequency, but by the model’s pre-trained term embeddings and fine-tuning on this particular task.

Transformer-based re-ranking models, although recent, are mainly used in single-turn passage re-ranking tasks~\cite{passagererankingbert, t5_reranking}. However, conversational search is more complex than simple passage ranking since we are in a multi-turn retrieval task. In this scenario, we have to not only manage the current query but also have a notion of the context where the query is posed. 
Therefore, while re-ranking, we want the model to consider the importance of the passage to the current query, as well as the importance of the passage given the conversational context. An example of this is present in Table~\ref{tab:example_conversation_dinossaur}. In turn 2, we want the model to focus on answering the query by retrieving passages that define ``extinction event'', but push to the top passages that discuss extinction events related to the conversational context, that in this case from the previous turn is the ``Cretaceous-Paleogene extinction event''.

\begin{table*}[ht]
\centering
\caption{Example of the need for conversational context to improve search results. The passages are adapted from the corresponding Wikipedia articles.}
\label{tab:example_conversation_dinossaur}
%\resizebox{\linewidth}{!}{%
\begin{tabular}{cp{0.2\linewidth}p{0.5\linewidth}}
\toprule
\textbf{Turn} &
  \textbf{Query} &
  \textbf{Passages} \\ \midrule
1 &
  What is a \ \underline{Tyrannosaurus}? &
  \underline{Tyrannosaurus} often called T-Rex, is one of the most well-represented of the large theropods and among the last non-avian dinosaurs to exist before the \underline{Cretaceous-Paleogene extinction event}. \\ \midrule
2 &
  What is an \underline{extinction event}? &
  An \underline{extinction-level event} is a widespread and rapid decrease in the biodiversity. An example of this event, the \underline{Cretaceous-Paleogene}, finished with 75\% of all species extinct. \\ \bottomrule
\end{tabular}%
%}
\end{table*}

Adding to this, the conversations can be long, spanning multiple turns, wherein each one there are various passages. This makes the approach of simply concatenating the multiple rounds together infeasible because of the maximum input of the BERT model (512 tokens), as well as the dispersion of topics throughout the conversation (topic changes) that can incorrectly influence the model, showing the need for specific models to tackle the conversational passage re-ranking task.

\subsubsection{\textbf{ConvBERT RNN}}
\label{convbert_rnn}
RNNs are generally used to model sequences of data and have been applied in various domains because of their ability to analyze the current input conditioned on previously seen inputs by the use of a hidden state~\cite{Hochreiter1997LongSM, gruOriginal, hredConversational}. In ConvBERT RNN, we aim to combine this architecture with the generated BERT embeddings and utilize the RNN's hidden state to model the conversational context between turns.

In ConvBERT RNN, we use a hierarchical structure inspired by the HRED~\cite{hredConversational} model, but we replace the bottom RNN (sentence level) with a BERT model that generates the sentence embeddings. This way, we can leverage the better sentence representations given by BERT and use an RNN to maintain the conversational context.

\begin{figure}[t]
  \centering
    {\includegraphics[width=1.0\linewidth]{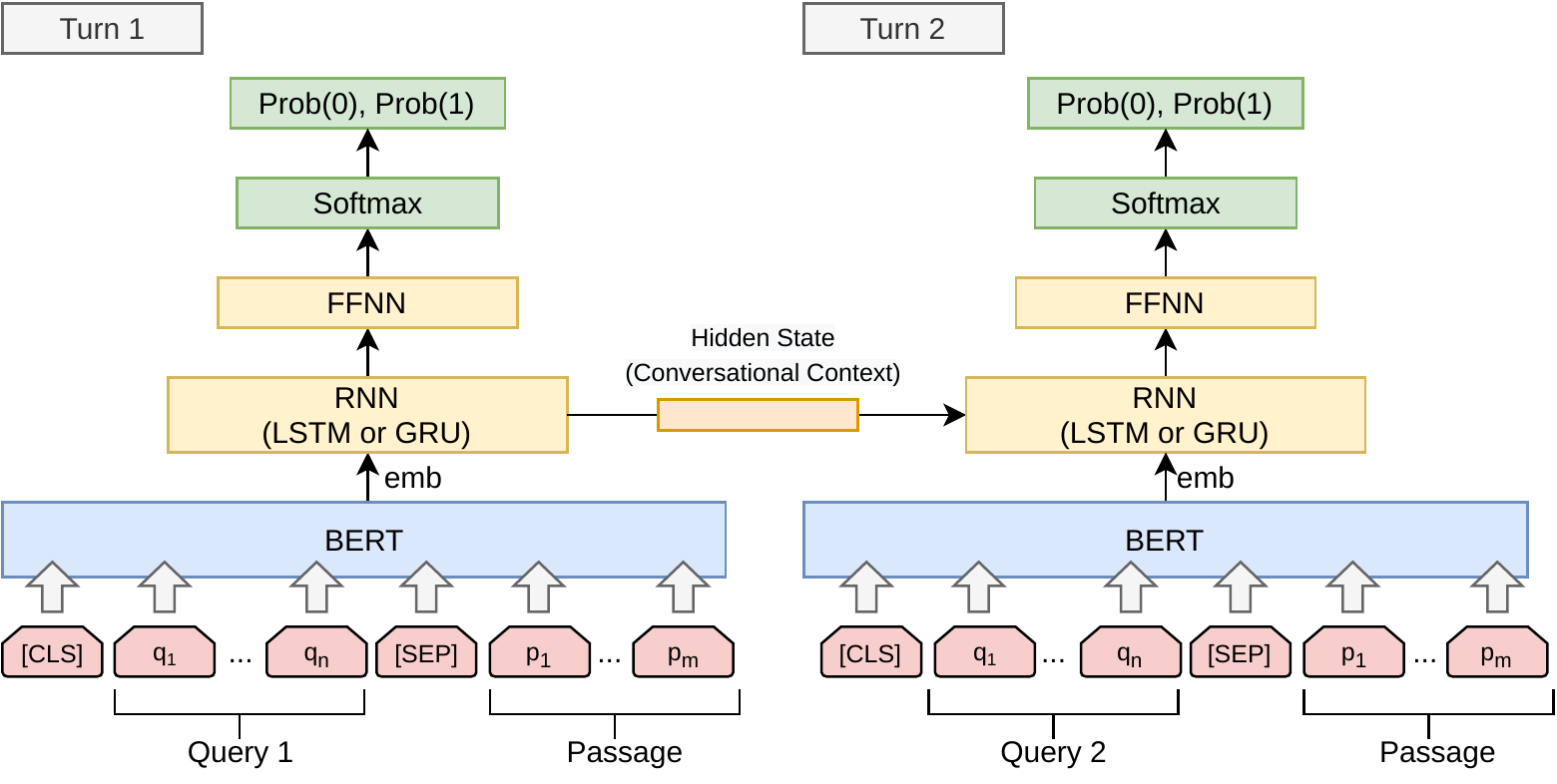}}
  \caption{ConvBERT RNN architecture. The left side of the figure represents the first turn in the conversation and the right side represents the second turn.}
  \label{fig:cov_bert_rnn_architecture}
\end{figure}

The ConvBERT RNN architecture can be seen in Figure~\ref{fig:cov_bert_rnn_architecture}, which represents two turns of information seeking. In the first turn, the query and each of the retrieved passages is encoded by BERT using the same input structure as in \cite{passagererankingbert} for single-turn passage ranking. 
In specific the input to BERT given a sequence of N tokens is given as:
\begin{equation}
\label{eq:embedding_bert}
    emb = BERT(``[CLS]\ q \ [SEP] \ p"),
\end{equation}
\noindent
where $emb \in \mathbb{R}^{N \times H}$ ($H$ is BERT embedding's size) is the embeddings matrix of all tokens, and \textit{[CLS]} and \textit{[SEP]} are special tokens in BERT's vocabulary, representing the classification and separation tokens, respectively.
From $emb$ we extract the embedding of the first token, which corresponds to the embedding of the  \textit{[CLS]} token in the last hidden state, $emb_{[CLS]} \in \mathbb{R}^{H}$. 
In theory, this token is able to represent the whole sequence since the model was fine-tuned on the relevance classification task using this embedding. This is also one of the most common ways of using BERT~\cite{bertOriginal, behaviourBert, passagererankingbert} in a re-ranking task.

After this, we pass BERT's \textit{[CLS]} embedding ($emb_{[CLS]}$) through an RNN such as, an LSTM~\cite{Hochreiter1997LongSM} or GRU~\cite{gruOriginal} to obtain the representation of the current query-passage pair in turn $i$ given the context ($out_{i}$), and the hidden state of the conversation ($h_{i}$), which will be passed to subsequent turns:
\begin{equation}
    out_{i}, \ h_{i} = RNN(emb_{[CLS]}, \ h_{i-1}),
\end{equation}
The representation of the query-passage pair ($out_i$) is subsequently used as input to a single-layer feed-forward neural network (FFNN) followed by a \textit{softmax} function to calculate the probability of the passage being relevant given a query $P(p|q)$: 
\begin{equation}
    P(p|q) = softmax(\ FFNN(out_i)\ ),
\end{equation}
being the passages ordered by this probability.
In the second turn, we apply the same procedure, but we also have access to the hidden state generated by the RNN in the previous turn ($h_{i-1}$).
We use this hidden state to maintain the conversational context by passing the hidden state generated by the first passage in the rank to the next turn. The rationale for this is that the most relevant passage, and the one that the user first sees, is the one that appears in the first position of the ranked list of results. In an interactive setting, we could replace this hidden state with the passage the user clicked, using the user's feedback instead of the model's previous output.

\subsubsection{\textbf{ConvBERT MemNet.}}
\label{conv_bert_mem_net}
A memory network can be used to mitigate the problems that arise from the use of RNNs, such as vanishing gradient problems~\cite{weston2014memory, sukhbaatar2015endtoend}. More important than that and specific to our task, is the problem that the previous queries and passages are always encoded in the RNN's hidden state, so previous queries and passages unrelated to the current query can add noise to the conversational context representation.
This problem is addressed by the memory networks because it explicitly stores the embeddings of each turn separately.
 
\begin{figure}[t]
  \centering
    {\includegraphics[width=0.98\linewidth]{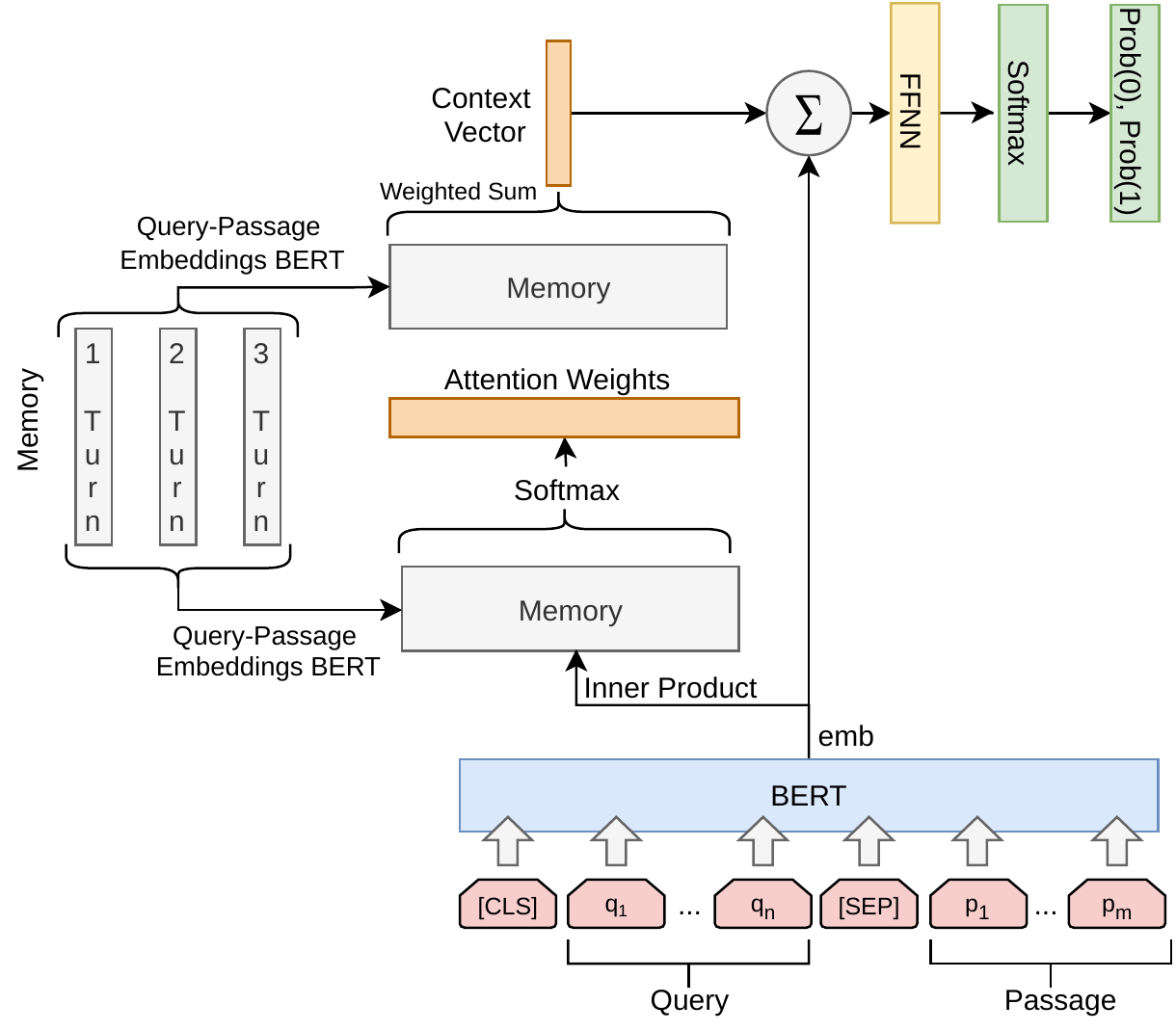}}
  \caption{ConvBERT MemNet architecture in the fourth turn of the conversation, storing the top query-passage embedding in each turn (3 previous turns in memory).}
  \label{fig:cov_bert_mem_net_architecture}
\end{figure}

Our approach with ConvBERT MemNet (Conversational BERT using Memory Networks) uses a single-layer, single-hop memory network~\cite{sukhbaatar2015endtoend} to model the conversational context. 
Figure~\ref{fig:cov_bert_mem_net_architecture} shows the implemented architecture in the fourth turn of the conversation, storing the top query-passage embedding in each turn (3 embeddings stored memories). 

The first step, and necessary in all turns of the conversation, is to use BERT to encode the query with each of the retrieved passages following the same input structure as ConvBERT RNN represented in equation~\ref{eq:embedding_bert}.

After getting the \textit{[CLS]} embeddings generated by the BERT model ($emb_{[CLS]}$), if we are in the first turn of the conversation, i.e., the memory is empty, we bypass the memory access and pass the embeddings directly to the feed-forward neural network followed by a \textit{softmax} function to obtain the probability of the passage being relevant. This is the same expression used in~\cite{passagererankingbert} because we are not in a conversational scenario:
\begin{equation}
    P(p|q) = softmax(\ FFNN(emb_{[CLS]})\ ).
\end{equation}

If we are not in the first turn (i.e., the memory is not empty), we follow the approach of Figure~\ref{fig:cov_bert_mem_net_architecture}. First, we perform the inner product between the current query-passage embedding, $emb_{[CLS]}$, and each of the memories (embeddings from previous turns), $m_i$, and then calculate the \textit{softmax} of the resulting inner products to get the attention weights, $a_i$, for each memory:
\begin{equation}
    a_i = softmax(emb_{[CLS]} \ m_i^T),
\end{equation}
\noindent
With these attention weights, we calculate a weighted sum with the memories, which returns the context vector $c$: 
\begin{equation}
    c = \sum_{i} \ a_i m_i,
\end{equation}
\noindent
Following that, we perform the sum of the weighted memories with the current embedding, and pass this through a linear layer and a \textit{softmax} function to get the probability of the passage being relevant:
\begin{equation}
    P(p|q)= softmax(\ FFNN(c+emb_{[CLS]})\ ).
\end{equation}

In this work, we store as memories the top passage given by the re-ranking algorithm in each turn. This is the same rationale used in ConvBERT RNN, but in the ConvBERT MemNet, the use of a dedicated memory provides a more flexible way of using the context. For example, instead of only using the embedding of the top passage retrieved as a memory, we can use several and adapt the weights given to each one depending on their rank. Also, in an interactive scenario, we could use as memories all of the passages previously clicked by the user, maintaining a complete history of the user's interactions with the system. 
We leave these ideas to future work.

With the ConvBERT MemNet architecture, we wish to mitigate the problems of the RNNs, while also having the advantage of keeping all of the previous sentence embeddings in memory in their original form, not losing any information. These embeddings can then be ``exploited'' or ``ignored'', depending on the attention weights given to each memory, so we believe that this architecture may be more robust to topic shifts. 

\section{Evaluation}
\label{evaluation_section}

\subsection{Datasets and Protocol}
\subsubsection{\textbf{CANARD Dataset}}~\cite{canYouUnpackIt}
We used this dataset to train and evaluate the T5 query rewriting model. This dataset contains conversational queries from the QuAC dataset~\cite{quac_dataset} and the respective rewrites performed by humans to form non-conversational queries. This dataset is composed of  31.538, 3.418, and 5.571 query-rewrites for the training, development, and test sets, respectively.

\subsubsection{\textbf{TREC CAsT Dataset}}~\cite{castoverview} 
This dataset was used to evaluate the performance of the retrieval system and to the best of our knowledge is the only dedicated conversational search dataset created. 
Considering the 2019 version as the training set, it contains in total 80 evaluation topics with an average of 10 turns. From these, 13 topics were annotated using a scale that ranges from 0 (not relevant) to 2 (highly relevant), and 20 topics were annotated on a scale that ranges from 0 (not relevant) to 4 (highly relevant). 
In this work, the 2020 version of this dataset is considered the evaluation set and contains 25 annotated topics using the same 0-4 scale previously mentioned. Also provided in the evaluation set are the queries rewritten (resolved) to include all of the information needed for the current turn, basically turning the task into a non-conversational task. These resolved queries can be used as an upper bound in terms of results when compared to the original (raw) conversational queries.
Adding to this, the track organizers also provide a query rewriting baseline (AUTO) based on the GPT-2 model. 

\subsubsection{\textbf{Experimental Protocol.}}
The performance of the query rewriting model T5 was evaluated using the BLEU-4~\cite{bleu} score between the model’s output and the queries rewritten by humans.

The passage retrieval and re-ranking components were evaluated using TREC CAsT's official metrics,  nDCG@3 (normalized Discounted Cumulative Gain at 3), MAP (Mean Average Precision), and MRR (Mean Reciprocal Rank), along with Recall.

\subsection{Implementation}

\subsubsection{\textbf{T5 Query Rewriting}}
\label{sub:eval_ query_rewriting}
We fine-tuned the T5 BASE model using standard maximum likelihood
on the CANARD dataset following~\cite{t5conversational}. In particular, we trained for 4000 steps using a learning rate of 0.0001, batches of 256 sequences, a maximum input sequence length of 512 tokens, a maximum output sequence length of 64 tokens, and greedy decoding.%a beam size of 1 (equivalent to no beam search). 

\subsubsection{\textbf{Indexing and Retrieval.}}
\label{indexing_retrieval_model}
To search and index, we used Anserini's~\cite{anserini} Python implementation Pyserini\footnote{\url{https://github.com/castorini/pyserini}}. 
We removed stop words using Lucene’s default list and applied the KStem\footnote{\url{ http://lexicalresearch.com/kstem-doc.txt}} stemming algorithm.
We tested with: BM25~\cite{bm25}, and language models with Jelinek-Mercer (LMJM) and Dirichlet (LMD) smoothing~\cite{languagemodelsmoothing}. From our preliminary experiments, LMD showed better results. This is consistent with previous work \cite{languagemodelsmoothing} that indicated that LMD works best with shorter queries which are common in a conversational search scenario. 
For this reason, LMD was the model used in all experiments.

\subsubsection{\textbf{ConvBERT Models.}}
\label{sub:eval_conv_bert}

\paragraph{\textbf{BERT Model and Fine-tuning}}
We experimented with the BASE version of BERT using the Transformers library from Huggingface~\cite{huggingface}. 
In particular, we used the fine-tuned BERT model \textit{nboost}\footnote{\url{https://huggingface.co/nboost}}, which has the same structure as the model in \cite{passagererankingbert} and was trained using the same method and data on MS MARCO dataset~\cite{marcoDataset} composed of 12.8 million query-passage pairs.

\paragraph{\textbf{ConvBERT Training}}
\label{sub:conv_bert_model_training}
Since the fine-tuned BERT model from \cite{passagererankingbert}, described in the previous paragraph already achieves a good performance in the single-turn relevance classification task, and because the amounts of data needed to fine-tune a BERT model in conversational data is not available, we freeze the fine-tuned model and use the generated embeddings as the basis for the ConvBERT models.
Therefore, we only train the top part of the ConvBERT models (RNN or Memory Network) using a binary conversational relevance classification task following~\cite{passagererankingbert} using the cross-entropy loss defined as:
\begin{equation}
    L = - \sum_{j \in J_{pos}} log(s_j) - \sum_{j \in J_{neg}} log(1-s_j),
\end{equation}
where $J_{pos}$ are the indexes of the relevant passages, $J_{neg}$ are the indexes of non-relevant passages, and $s_j$ is the score given by the ConvBERT model to the query-passage pair.

\paragraph{\textbf{Data Preprocessing and Training Procedure}}
\label{data_preprocessing}
Regarding the data, to train the \textit{ConvBERT} models, we need conversations composed of conversational queries about a specific topic with corresponding passages labeled as relevant and non-relevant to the query.
The input to the model in each turn will then be a query-passage pair, and the output is the probability of the passage being relevant to the query, which in the re-ranking task is used to order the passages.
In specific, we used TREC CAsT 2019. To use this data in a binary classification task we simplified the annotations to only 0 and 1 by considering the annotations with 1 and 2 as 1 in the 0-2 scale, and by considering the annotation with 3 and 4 as 1 in the 0-4 scale, creating a dataset where 80\% of the passages have label 0.
After this pre-processing, for each topic, we create approximately $X$ conversations by randomly sampling without replacement an annotated passage for each turn (query), where $X$ is the number of judged passages in the first turn. This approach creates 2780 conversations for training, with an average of 10 turns.
With this set of conversations, we considered a batch to be a full conversation topic to guarantee that the model sees a full conversation in succession, instead of parts of different conversations at each gradient calculation step.

To validate our models and choose the parameters, we used a train-validation split of 75\%, 25\%, using 5-fold cross-validation over the full training set. Since we are working with conversational topics, when splitting the dataset, we enforce that the same topic only appears in one of the sets.
With respect to the hyperparameters, we tested the batch sizes [1, 2, 4] conversations (each conversation has on average 10 query-passage pairs), and learning rates [0.001, 0.0001] using the Adam optimizer~\cite{kingma2017adam}.
We then chose the parameters that achieved the highest average validation F1 score considering all folds.

\subsection{Methods}
With respect to the queries used, we consider the following methods:
\begin{itemize}
    \item \textbf{Raw} – original conversational queries (lower-bound);
    \item \textbf{Manual} – queries manually rewritten by the track organizers (upper-bound);
    \item \textbf{Auto} – queries rewritten provided by the track organizers obtained with a GPT-2 model~\cite{gpt2_query_rewriting};
    \item \textbf{Prefix+CorefPronoun} - queries rewritten using AllenNLP with the first query of the conversation prefixed; 
    \item \textbf{T5} – queries rewritten using the T5 model;
    \item \textbf{T5+Union} – Union of the queries resolved with T5.
\end{itemize}

In retrieval, we always use LMD. Regarding the re-ranking models we considered:
\begin{itemize}
    \item \textbf{BERT BASE} - This is a baseline model composed of a BERT model with a linear layer on top from \cite{passagererankingbert};
    \item \textbf{BERT Linear} - Has the same structure as the previous, but the linear layer is only trained on CAsT's 2019 dataset. This model serves as a baseline to compare the effectiveness of the architectures proposed in Section \ref{conv_bert};
    \item \textbf{ConvBERT RNN} - Model described in Section \ref{convbert_rnn}, consisting of a BERT model with an RNN on top of one of these types: LSTM, GRU, BiLSTM, and BiGRU. The RNN is trained on CAsT 2019 dataset and in the case of the bidirectional RNNs the representations from both directions are concatenated together;
    \item \textbf{ConvBERT MemNet} - BERT model with a Memory Network on top (Section \ref{conv_bert_mem_net}) trained on CAsT 2019 dataset.
\end{itemize}
As stated before, all re-ranking methods use the embeddings generated by a BERT model fine-tuned on MS MARCO following~\cite{passagererankingbert}.

\subsection{T5 Query Rewriting Results}
Table~\ref{table:t5_bleu_results} shows the BLEU-4 scores in CANARD and in TREC CAsT 2019 \textit{Manual} queries. 
The first two rows are from the paper that introduced CANARD~\cite{canYouUnpackIt}. The T5-BASE row is from \cite{t5conversational}, and the last row corresponds to our implementation (Section \ref{sub_sub_t5}). Our results are lower in CANARD and higher in TREC CAsT when compared to \cite{t5conversational}. We attribute this to the construction of the input sequence since the exact method is not specified in \cite{t5conversational}.

From the results obtained and from direct observation of the outputs of the model, we concluded that the model is performing both explicit and implicit coreference resolution, being this the key advantage of this query rewriting model over basic coreference resolution like the one presented in Section~\ref{sub_sub_allennlp}.

\begin{table}[t]
\centering
\caption{BLEU-4 scores for CANARD dev and test sets and for TREC CAsT 2019 annotated resolved queries (\textit{Manual}).}
\label{table:t5_bleu_results}
\begin{tabular}{lccc}
\toprule
{ }           & \multicolumn{2}{c}{{ \textbf{CANARD}}}  & { \textbf{CAsT 2019}} \\ \cline{2-4} 
{ } &
  { \textbf{Dev}} &
  { \textbf{Test}} &
  { \textbf{Manual Queries}} \\ \midrule
{ Human~\cite{canYouUnpackIt}} & \multicolumn{2}{c}{{ 59.92}}            & { -}                  \\
{ Raw~\cite{canYouUnpackIt}}   & { 33.84} & { 47.44} & { -}                  \\
{ T5-BASE~\cite{t5conversational}} &
  { \textbf{59.13}} &
  { \textbf{58.08}} &
  { 75.07} \\ \midrule
\multicolumn{4}{l}{\textit{Our implementation:}}                                                                                           \\
{ T5 (Section \ref{sub_sub_t5})}      & { 58.48} & { 56.84} & { \textbf{79.67}}     \\ \bottomrule
\end{tabular}%
\end{table}

\subsection{ConvBERT Results}
Table~\ref{tab_retrieval_results} shows the results of the models trained on CAsT, ConvBERT architectures and \textit{Linear}, as well as the BERT \textit{BASE} fine-tuned on MS MARCO. These last two models do not have a notion of context. All methods re-rank the top-1000 passages retrieved with LMD. 
The underlines indicate the best method in each group of queries, and the bold results indicate the best method overall, excluding manually resolved queries that are always the upper bound of the experiment.
When re-ranking, we use the same queries at retrieval and re-ranking phases, however, because \textit{T5+Union} issues multiple queries at re-ranking time we only use the query generated by T5 for the last turn. 
As a final note, when no re-ranker is specified this indicates the absence of a re-ranking step (only retrieval is performed).

\begin{table}[htbh]
\centering
\caption{Results on CAsT 2020 evaluation set using LMD and re-ranking the top-1000 passages with different models.}
\label{tab_retrieval_results}
\resizebox{\linewidth}{!}{%
\begin{tabular}{llcccc}
\toprule
\multicolumn{1}{c}{\textbf{Queries}} & \multicolumn{1}{c}{\textbf{Re-ranking}} & \textbf{Recall} & \textbf{MAP} & \textbf{MRR} & \textbf{nDCG@3} \\ \midrule
Raw                 & -       & 0.296 & 0.065          & 0.187          & 0.078          \\
Raw                 & Linear  & 0.296 & 0.109          & 0.300          & 0.158          \\
Raw                 & GRU     & 0.296 & 0.116          & 0.327          & 0.178          \\
Raw                 & LSTM    & 0.296 & 0.116          & 0.334          & 0.185          \\
Raw                 & Bi-GRU  & 0.296 & 0.114          & 0.324          & 0.179          \\
Raw                 & Bi-LSTM & 0.296 & \underline{0.119} & \underline{0.341} & \underline{0.190} \\
Raw                 & MemNet  & 0.296 & 0.113          & 0.311          & 0.174          \\
Raw                 & BASE    & 0.296 & 0.118          & 0.339          & 0.186          \\ \hline
AUTO                & -       & 0.551 & 0.129          & 0.392          & 0.170          \\
AUTO                & Linear  & 0.551 & 0.202          & 0.512          & 0.290          \\
AUTO                & GRU     & 0.551 & 0.221          & 0.557          & 0.321          \\
AUTO                & LSTM    & 0.551 & 0.224          & 0.574          & 0.331          \\
AUTO                & Bi-GRU  & 0.551 & 0.220          & 0.559          & 0.321          \\
AUTO                & Bi-LSTM & 0.551 & 0.228          & \textbf{0.577} & \underline{0.324} \\
AUTO                & MemNet  & 0.551 & 0.209          & 0.532          & 0.310          \\
AUTO                & BASE    & 0.551 & \textbf{0.229} & 0.570          & 0.321          \\ \hline
Prefix+CorefPronoun & -       & 0.515 & 0.108          & 0.326          & 0.128          \\
Prefix+CorefPronoun & Linear  & 0.515 & 0.167          & 0.454          & 0.243          \\
Prefix+CorefPronoun & GRU     & 0.515 & 0.178          & 0.472          & 0.254          \\
Prefix+CorefPronoun & LSTM    & 0.515 & 0.178          & 0.471          & 0.255          \\
Prefix+CorefPronoun & Bi-GRU  & 0.515 & 0.177          & \underline{0.478} & 0.256          \\
Prefix+CorefPronoun & Bi-LSTM & 0.515 & 0.181          & \underline{0.478} & 0.257          \\
Prefix+CorefPronoun & MemNet  & 0.515 & 0.173          & 0.468          & 0.248          \\
Prefix+CorefPronoun & BASE    & 0.515 & \underline{0.184} & 0.470          & \underline{0.260} \\ \hline
T5                  & -       & 0.529 & 0.124          & 0.361          & 0.158          \\
T5                  & Linear  & 0.529 & 0.193          & 0.505          & 0.277          \\
T5                  & GRU     & 0.529 & 0.209          & 0.534          & 0.308          \\
T5                  & LSTM    & 0.529 & 0.210          & 0.532          & 0.306          \\
T5                  & Bi-GRU  & 0.529 & 0.207          & 0.542          & 0.312          \\
T5                  & Bi-LSTM & 0.529 & 0.214          & 0.537          & 0.315          \\
T5                  & MemNet  & 0.529 & 0.198          & 0.518          & 0.285          \\
T5                  & BASE    & 0.529 & \underline{0.220} & \underline{0.557} & \underline{0.320} \\ \hline
T5+UNION            & -       & 0.552 & 0.109          & 0.348          & 0.133          \\
T5+UNION            & Linear  & 0.552 & 0.198          & 0.516          & 0.288          \\
T5+UNION            & GRU     & 0.552 & 0.213          & 0.550          & 0.318          \\
T5+UNION            & LSTM    & 0.552 & 0.213          & 0.552          & 0.313          \\
T5+UNION            & Bi-GRU  & 0.552 & 0.211          & 0.551          & 0.320          \\
T5+UNION            & Bi-LSTM & 0.552 & 0.216          & 0.553          & 0.315          \\
T5+UNION            & MemNet  & 0.552 & 0.204          & 0.537          & 0.304          \\
T5+UNION            & BASE    & 0.552 & \underline{0.221} & \underline{0.575} & \textbf{0.330} \\ \hline
Manual              & -       & 0.762 & 0.213          & 0.586          & 0.284          \\
Manual              & Linear  & 0.762 & 0.327          & 0.736          & 0.440          \\
Manual              & GRU     & 0.762 & 0.358          & 0.786          & 0.495          \\
Manual              & LSTM    & 0.762 & 0.363          & 0.781          & 0.501          \\
Manual              & Bi-GRU  & 0.762 & 0.357          & 0.791          & 0.505          \\
Manual              & Bi-LSTM & 0.762 & \underline{0.368} & \underline{0.790} & \underline{0.509} \\
Manual              & MemNet  & 0.762 & 0.338          & 0.743          & 0.469          \\
Manual              & BASE    & 0.762 & \underline{0.368} & 0.778          & 0.499          \\ \bottomrule
\end{tabular}%
}
\end{table}

\subsubsection{\textbf{Query Rewriting Results Analysis}}
When evaluating the results of Table~\ref{tab_retrieval_results}, the first thing that becomes evident is the need for a query rewriting method, evidenced by the low scores in all metrics when using the \textit{Raw} queries. Regarding the query rewriting methods developed, we see that all of them had the desired effect, improving on the results of the original conversational queries (\textit{Raw}) by 64-117\% in nDCG@3, and in turn, moving closer to the upper bound corresponding to the manually rewritten queries (\textit{Manual}). 
Going in more detail, considering only methods without a re-ranking step, in terms of recall the best methods were \textit{AUTO} (0.551) and \textit{T5+UNION} (0.552). In particular, the \textit{T5+UNION} result shows that the Union approach of Section \ref{sub_sub_previous_queries} combined with the \textit{T5} query rewriting method can provide various relevant passages.
With respect to the other metrics without a re-ranking step, we observe that \textit{T5+UNION} is not the best method since the fusion of the lists by the retrieval score although providing more relevant passages, evidenced by the higher recall, is not the best in metrics that evaluate the earlier positions of the rank. So, without re-ranking, the best methods are \textit{AUTO} and \textit{T5}, achieving an nDCG@3 of 0.170 and 0.158, respectively. 

\subsubsection{\textbf{Re-ranking Models Results Analysis}}
Regarding the application of the re-ranking models in Table \ref{tab_retrieval_results}, we observe that they improve all metrics except for recall because it uses exactly the same set of passages in retrieval and re-ranking. We expected these improvements because of BERT's ability to judge if a passage is actually relevant to a particular query, not by term matching, but by having an understanding of the text in the query and passage.
Comparing the architectures trained on the CAsT dataset that make use of the conversational context (\textit{ConvBERT} architectures) with the baseline approach that uses a single linear layer on top of BERT's output trained on the same data (\textit{Linear}), we see that the use of context is indeed helpful in all query formulations. 
For example, in the \textit{AUTO} queries, we see that the nDCG@3 achieved by the \textit{ConvBERT} \textit{Bi-LSTM} model is 11.7\% higher than the one obtained with the \textit{Linear} model.
These results verify our hypothesis that the use of state, in the form of BERT embeddings stored in the hidden state of the RNN and in the memory of the Memory Network, are important and can be used in a conversational re-ranking scenario.

In further detail, when we compare the \textit{ConvBERT RNN} and \textit{ConvBERT MemNet} with each other, we see that in most metrics, more than one RNN-based approach surpassed the \textit{MemNet} architecture. 
We consider that this may happen because the conversations are not very long, on average 8 turns, and so the RNN's hidden state is able to keep the important information to judge the relevance of a passage.
Analyzing the different RNNs architectures, we see that there is not a great difference in the best models. 
Despite this, we generally see that the bidirectionality brings improvements to both \textit{GRU} and \textit{LSTM} models, and that in most query formulations the best model is the \textit{Bi-LSTM}. 

Regarding the different query types in the re-ranking architectures, we see that the best methods in most metrics are \textit{AUTO} and \textit{T5+UNION}. 
The results in \textit{T5+UNION} are in contrast with the results using only retrieval because the re-ranking model is able to successfully use the more varied set of passages provided by \textit{T5+UNION} to its advantage, thanks to its better relevance assessment when compared to the scores given by the LMD retrieval model. The results also show that the developed T5-based methods have a similar performance to the GPT-2 model used to create the \textit{AUTO} queries.

To finalize, comparing the results of the \textit{ConvBERT} architectures to the BERT \textit{BASE} model fine-tuned on MS MARCO, we see that the results are close to each other despite the large difference in the amount of training data.
In more detail, the top part of the \textit{ConvBERT} architectures was trained on CAsT using approximately 29270 conversational query-passage pairs, while the linear layer in \textit{BASE} method was trained on MS MARCO on 12.8 million query-passage pairs~\cite{passagererankingbert}. 
This result further shows the typical paradigm present in Transformer-based methods that having more data available to train results in a better model even when the model is trained on different data~\cite{t5}.
This also shows that the embeddings generated by BERT are of high quality, being able to create a comparable model (i.e. \textit{ConvBERT}) even with limited training data.

\subsection{Analysis of Conversational Patterns}
\label{sec:final_architecture}

\subsubsection{\textbf{Per-Turn results analysis}}
Due to the conversational aspects of the task, it is important to evaluate the performance of the different models with respect to the turn depth.

\begin{figure}[t]
  \centering
    {\includegraphics[width=0.85\linewidth, trim=5pt 8pt 5pt 5pt, clip]{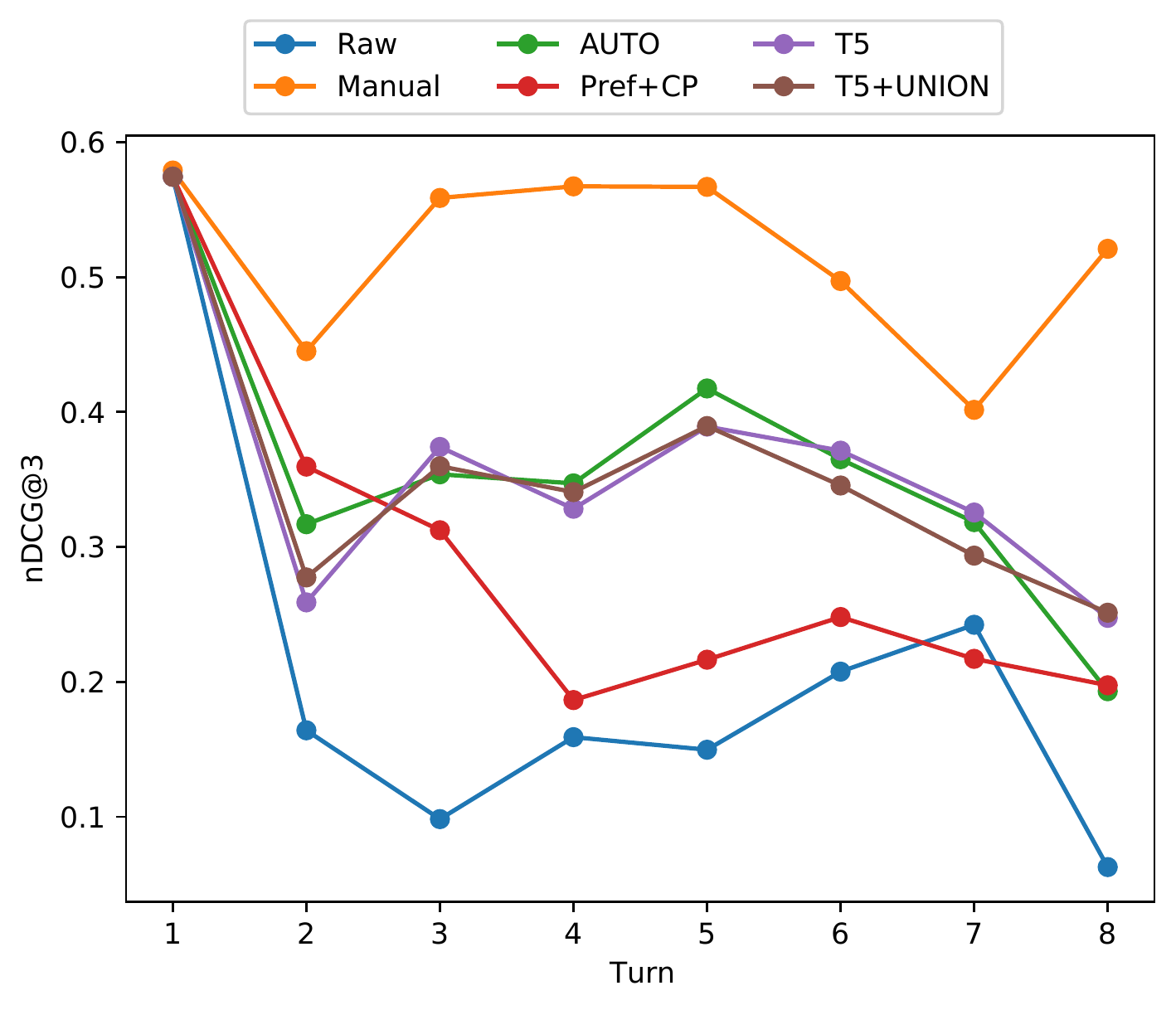}}
      \caption{nDCG@3 by turn depth using various query types, using the \textit{ConvBERT Bi-LSTM} model as re-ranker.}
  \label{fig:turn_depth_by_query}
\end{figure}

\paragraph{\textbf{Query Types By Turn Depth}}
Figure~\ref{fig:turn_depth_by_query} shows the nDCG@3, the main metric of TREC CAsT, by turn depth obtained with different query rewriting methods using the best performing ConvBERT model, \textit{ConvBERT Bi-LSTM}, as the re-ranker.
As expected, the \textit{Raw} queries achieved good results in the first turn since they are not conversational, but after this results suffer a large decline due to the introduction of context from previous turns. 
The \textit{Manual} queries, also as expected, are the best-performing ones because they turn this task into a typical information retrieval task and so are not directly affected by turn depth.
\textit{Pref\-+CorefPronoun} is better than \textit{Raw} in most turns although a degradation in performance is visible due to failed coreference resolution and because of the noise introduced by the first query.
With respect to the approaches that use large pre-train models (\textit{AUTO}, \textit{T5} and \textit{T5+UNION}), we see that all of them have a similar behaviour showing that the different methods have comparable results in most turns.
These results also demonstrate the importance of creating good query rewriting methods so that the performance does not decrease the further we are in the conversation.

\begin{figure}[t]
  \centering
  {\includegraphics[width=0.87\linewidth, trim=5pt 8pt 5pt 5pt, clip]{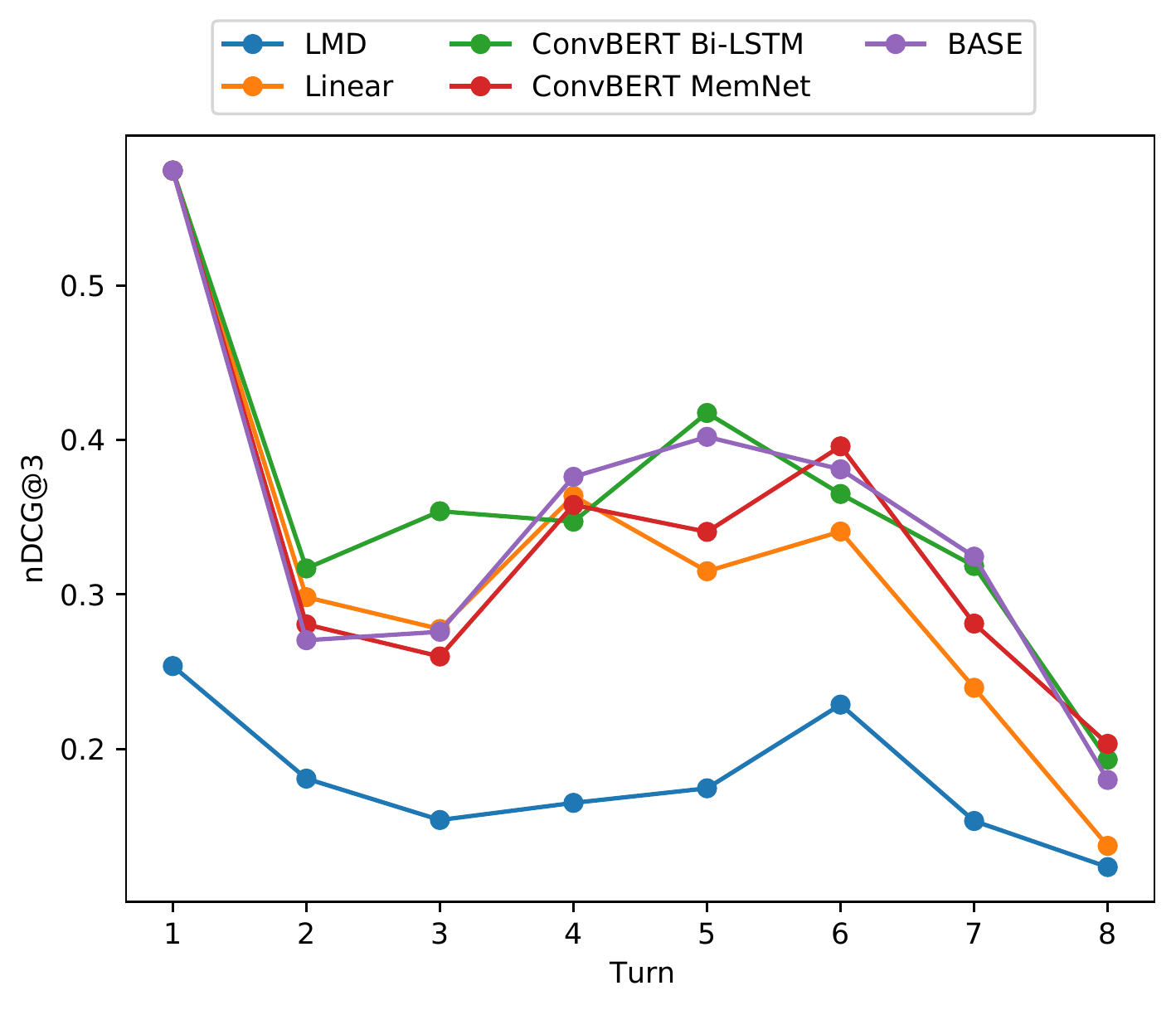}}% 
  \caption{nDCG@3 by turn depth using various re-ranking models using the \textit{AUTO} queries.}
  \label{fig:turn_depth_by_model}
\end{figure}

\paragraph{\textbf{Re-ranking Models By Turn Depth}}
Figure~\ref{fig:turn_depth_by_model} shows the nDCG@3 by turn depth using \textit{AUTO} queries, using different re-ranking models. 
As seen before, re-ranking has a big influence on performance with LMD (no re-ranking), achieving lower scores by a large margin in all turns. The other models present similar behaviours between themselves since all of them share the same BERT model.
Confirming our previous results, the \textit{Linear} model has the worst performance among the re-ranker in most turns because it has no notion of context. This seems to be particularly visible from turn 5 onwards.
BERT \textit{BASE} and \textit{ConvBERT} \textit{Bi-LSTM} have similar performance across the conversational turns, although in turn 2 and 3 we observe a greater difference between both models with the \textit{ConvBERT Bi-LSTM} re-ranker outperforming the model without context \textit{BASE}.
\textit{MemNet} generally has lower results than \textit{ConvBERT Bi-LSTM} in most conversational turns, although an increase in performance is observed in turn 6.

\subsubsection{\textbf{Memory Network Attention}}
\begin{figure}[t]
  \centering
    {\includegraphics[width=0.95\linewidth]{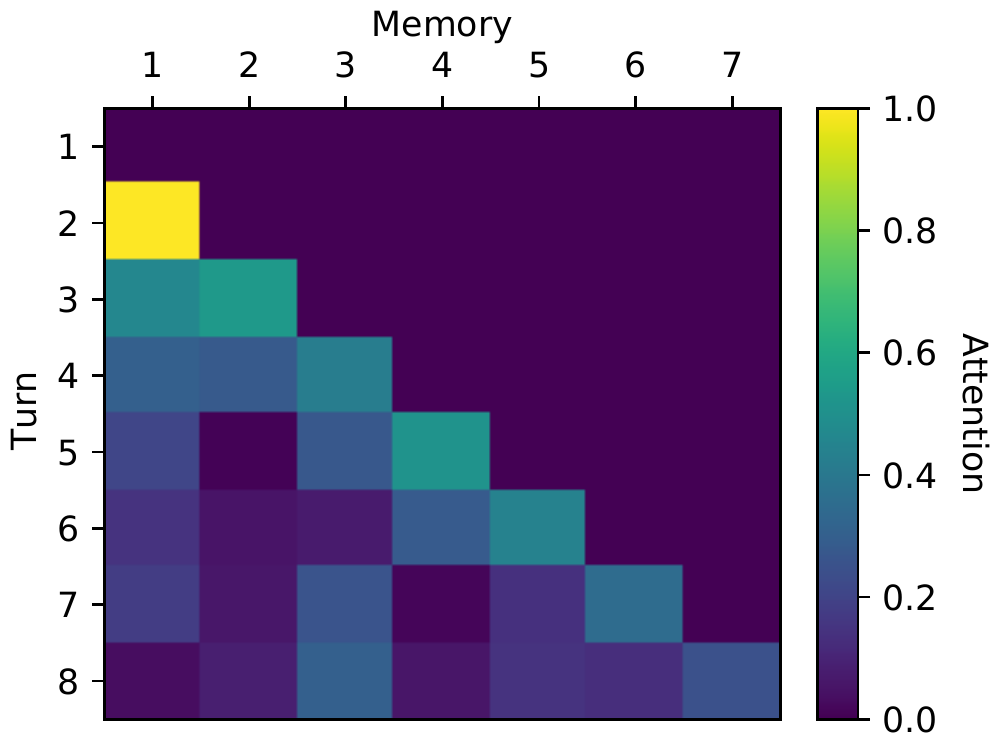}}
  \caption{Memory Network attention by turn depth.}
  \label{fig:mem_net_attention}
\end{figure}

To understand more about the behaviour of the memory network in \textit{ConvBERT MemNet}, Figure \ref{fig:mem_net_attention} shows the average attention weights given to each memory by turn depth using the \textit{Manual} queries. 
As expected, a large portion of the attention is given to the immediate previous turn, which indicates that the current turn is more related to the previous as in a normal conversation, where we reference more recent knowledge. Another interesting behaviour demonstrated is that some attention is also given to the first turn even in advanced turns of the conversation, which confirms our hypothesis that the first turn is important to set the topic of the conversation.

\subsubsection{\textbf{BERT Embeddings}}
\begin{figure}[t]
  \centering
    {\includegraphics[width=0.98\linewidth, trim=5pt 10pt 5pt 5pt, clip]{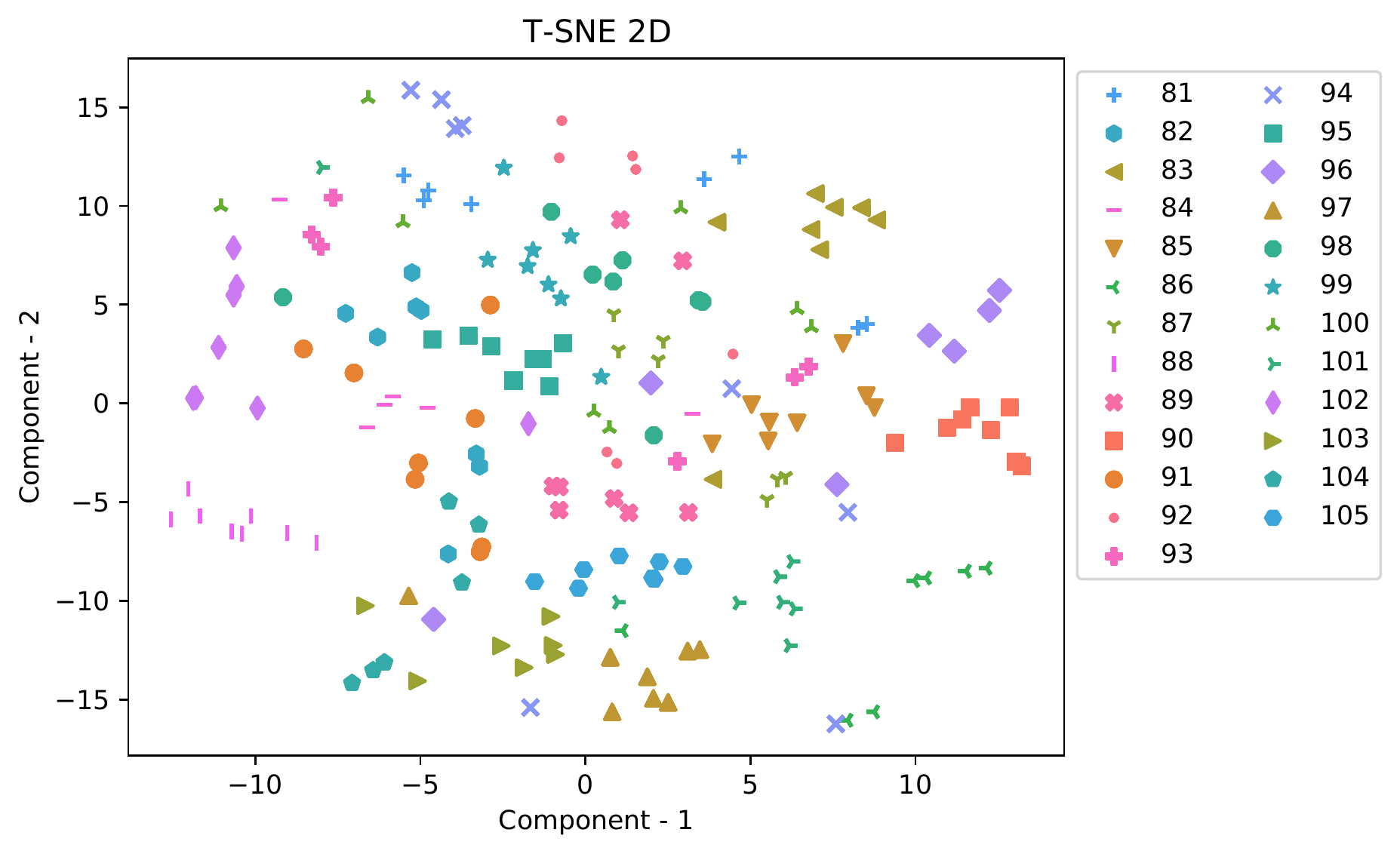}}
  \caption{BERT BASE embeddings by topic using t-SNE.}
  \label{fig:t_sne_bert_by_topic}
\end{figure}

In Figure \ref{fig:t_sne_bert_by_topic}, we show the embeddings generated by the BERT model fine-tuned on MS MARCO~\cite{passagererankingbert} using the \textit{Manual} queries and respective top-ranked passage for each turn, projected from a 768 positions vector to a 2D plane using the t-distributed Stochastic Neighbour Embedding (t-SNE) algorithm~\cite{tsne_paper}. 
We can see that the embeddings generated by the model for the same topic generally form clusters. This is very insightful because it shows that the embeddings not only carry the information necessary to perform the re-ranking, but also have a similar structure when in the same topic. This result is very promising and shows that the rationale of using these embeddings in conversational search is also valid in the embedding space. 

\section{Conclusions}
\label{conclusions_section}
In this paper, we explored the context present in previous utterances, as well as in the embeddings generated by a BERT model in the conversational search task. The key findings of our work are the following:
\begin{itemize}
    \item \textbf{Context in previous utterances} - Rewriting the user utterance using the conversation's context is critical. We tested different query rewriting approaches and from the developed methods T5 delivered the most significant boost in performance when combined with the previous queries in the conversation.
    \item \textbf{BERT embeddings can capture conversational context} - The proposed context-aware re-ranking models showed that modeling the sequence of BERT embeddings throughout the conversation results in an improvement of 11.7\% over a linear model trained on the same data.
\end{itemize}

As future work, we plan to develop new query rewriting methods that take into account the context present in the passages retrieved. 
Other research paths include adapting existing conversational question-answering datasets to the conversational search task in order to obtain more training data for the \textit{ConvBERT} re-rankers. Finally, thanks to the modularity of our approach, we can also replace the BERT model by another more recent Transformer-based model.

\section*{Acknowledgment}
This work has been partially funded by the iFetch project, Ref. 45920 co-financed by ERDF, COMPETE 2020, NORTE 2020, the CMU Portugal project GoLocal Ref. CMUP-ERI/TIC/0046/2014 and by the project NOVA LINCS Ref. UID/CEC/04516/2013.

\ifCLASSOPTIONcaptionsoff
  \newpage
\fi

\end{document}